\documentclass[%
 reprint,
groupedaddress,
 amsmath,amssymb,
 aps,showkeys
pra,
]{revtex4-2}
\usepackage{pdfpages}
\usepackage{pgffor} % for loops
\usepackage{graphicx}% Include figure files
\usepackage{dcolumn}% Align table columns on decimal point
\usepackage{bm}% bold math
\makeatletter
\AtBeginDocument{\let\LS@rot\@undefined}
\makeatother

\begin{document}

%\preprint{APS/123-QED}

\title{Coupled micro-resonators for second-order integrated nonlinear optics}% Force line breaks with \\
%\thanks{A footnote to the article title}%

\author{Yannick Dumeige}
%\altaffiliation[Also at ]{Physics Department, XYZ University.}%Lines break automatically or can be forced with \\
%\author{Second Author}%
 \email{yannick.dumeige@univ-rennes.fr}
\affiliation{%
 Université de Rennes, ENSSAT, CNRS, Institut FOTON-UMR 6082, 6 rue de Kerampont, 22300 Lannion, France
 %This line break forced with \textbackslash\textbackslash
}%

%\collaboration{MUSO Collaboration}%\noaffiliation

\author{Yoan Léger}
% \homepage{http://www.Second.institution.edu/~Charlie.Author}
\affiliation{
 Université de Rennes, INSA Rennes, CNRS, Institut FOTON-UMR 6082, F-35000 Rennes, France
% This line break forced% with \\
}%
%\affiliation{
% Third institution, the second for Charlie Author
%}%
%\author{Delta Author}
%\affiliation{%
% Authors' institution and/or address\\
% This line break forced with \textbackslash\textbackslash
%}%

%\collaboration{CLEO Collaboration}%\noaffiliation

\date{\today}% It is always \today, today,
             %  but any date may be explicitly specified

\begin{abstract}
We investigate second-order nonlinear processes in a system of two coupled identical optical micro-resonators. The double resonance and phase-matching conditions are simultaneously obtained thanks to the frequency splitting induced by the resonator coupling. The analysis made in the framework of the coupled mode theory is applied to the second harmonic generation process in two whispering gallery mode microdisks made in III-V materials.
\end{abstract}

%Use showkeys class option if keyword
                              %display desired
\maketitle

%\tableofcontents

Optical micro-resonators are of great interest for nonlinear photonics since they enable simultaneously, high integration, low power operation and high stability \cite{Lin:17}. One of the most striking applications based on third order nonlinear optics in micro-resonators is the integration of frequency comb sources on a chip \cite{Stern:18}. second-order nonlinear optics can also take benefit from the use of optical micro-resonators which could lead to the miniaturization of frequency converters, optical parametric oscillators or entangled photon sources \cite{Liu:23}. In this aim, the platform of III-V semiconductors is promising since they have high second-order susceptibilities, high refractive index and direct band gap which let envisage micro- or nano-photonics nonlinear or quantum active devices \cite{Baboux:23} in a wide spectral band. Considering GaP for instance, the transparency domain ranges from $0.55~\mu\mathrm{m}$ to $11~\mu\mathrm{m}$ \cite{Wang:24} up to wavelength inaccessible to lithium niobate. III-V materials are isotropic, thus birefringence phase matching  is not possible and several methods have been demonstrated to circumvent this issue \cite{Fiore:98, Dumeige:02, Ducci:04, Helmy:06, Pantzas:22}. In this context, natural quasi-phase matching in semiconductor whispering gallery mode (WGM) resonators have been proposed \cite{Dumeige:06a, Lorenzo_Ruiz:20} and experimentally demonstrated \cite{Kuo:14}. Nevertheless, due to the very high dispersion of III-V semiconductors, practical implementations of this quasi-phase matching scheme also rely on the use of modal phase matching \cite{Mariani:14, Lake:16, Chang:19} reducing the effective nonlinear susceptibility due to a low interacting fields overlap. Furthermore, complex numerical techniques can be implemented to optimize the phase matching and the nonlinear overlap in the design of doubly resonant nonlinear photonic structures \cite{Simonneau:97, Bi:12, Lin:16}. From another point of view, it has been shown that the coupling of several resonators or of counter-propagating modes within a single resonator allows the control or the engineering of the dispersion in $\chi^{(3)}$ processes \cite{Armaroli:15,Lu:23,Swarnava:25}. Coupled resonator structures can be used to reach functionalities which are not attainable with single nonlinear resonators \cite{Dumeige:06b}. If we restrict ourselves to the case of second-order effects, only few works have been devoted to the coupling of nonlinear micro-resonators \cite{Xu:00, Dumeige:07b, Dumeige:11}. In particular, it has been shown in the tight binding approximation that the dispersion induced by the weak coupling of resonators can be used to reach the phase matching condition for SHG in a long chain of $\chi^{(2)}$ microcavities \cite{Xu:00}. 

In this paper, we propose a simple and comprehensive design method using the unique properties of the coupling between two identical WGM resonators to obtain the phase matching condition. Using a benchmark of GaP WGM microdisks, we show that the strong coupling between two resonators helps to reach the natural quasi-phase matching condition with low radial order modes in the process of second harmonic generation (SHG). This approach would lead to reach the low-power high-conversion regime on a chip.

\begin{figure}[ht]
\centering
\includegraphics[width=8.5cm]{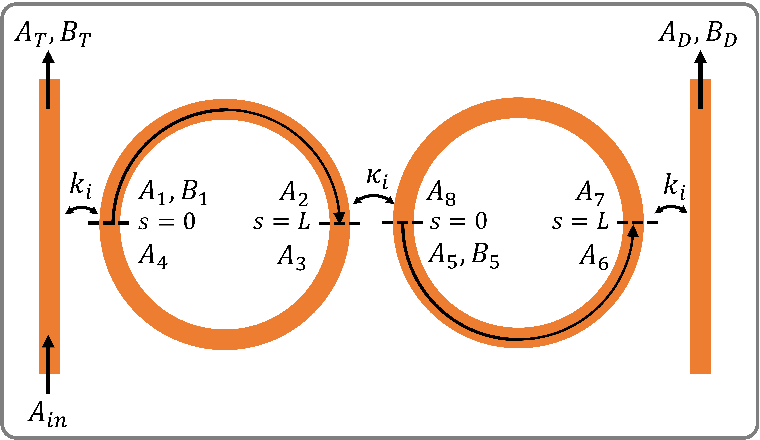}
\caption{System of coupled ring or disk resonators of perimeter $2L$. $A_i$ and $B_i$ are the F and SH mode amplitudes within the structure. $k_i$ are the coupling between the access bus waveguides and the structure. $\kappa_i$ are the coupling coefficient between the two resonators. $A_{in}$ is the input F mode amplitude. $A_T$ and $B_T$ are the F and SH output mode amplitude on the trough port. $A_D$ and $B_D$ are the output mode amplitude on the drop port.}
\label{Fig1}
\end{figure}

\begin{figure}[h!]
\centering
\includegraphics[width=6.5cm]{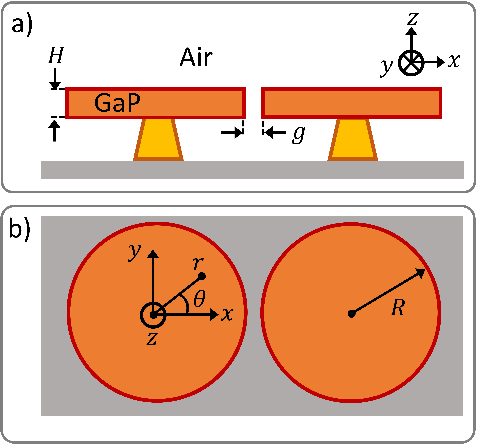}
\caption{Description of the structure numerically studied consisting of two GaP WGM microdisks embedded in air. a) side view and b) top view. $g$ is the air gap separating the two disks, $H$ and $R$ are their thickness and radius.}
\label{Disks}
\end{figure}

\begin{figure}[h!]
\centering
\includegraphics[width=\linewidth]{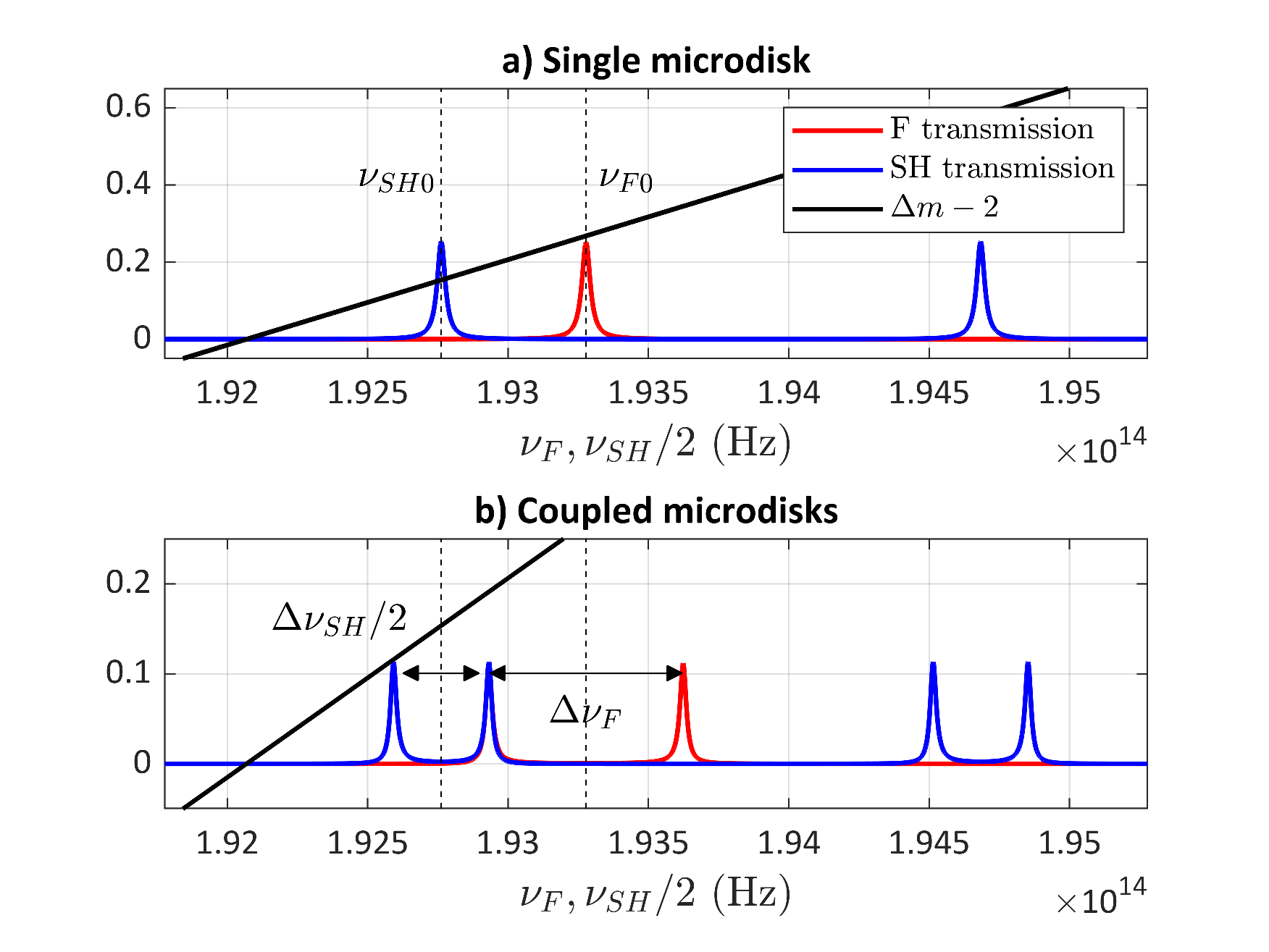}
\caption{Transmission spectra calculated on the drop port and azimutal quantum number mismatch. a) for a single resonator the double resonance $\nu_{F0}=\nu_{SH0}/2$ and the condition $|\Delta m-2|<1/2$ can not be fulfilled due to the dispersion. b) By coupling the two disks it is possible to reach both conditions simultaneously.}
\label{Spectrelin}
\end{figure}

Figure \ref{Fig1} is a sketch of the structure studied in this paper. It consists of two rings or disks of radius $R=L/\pi$ mutually coupled due to the overlap of the evanescent part of their modes. The resonators are coupled to two insertion and extraction waveguides with the same rate. The curvilinear abscissa along the resonator is called $s$. The fundamental (F) fields are denoted $A_i(s)=\mathcal{A}_i(s)e^{j(\omega t-\beta_1s)}$ and the second harmonic (SH) fields $B_i(s)=\mathcal{B}_i(s)e^{j(2\omega t-\beta_2s)}$ with $i\in[1,8]_{\mathbb{N}}$ and where $\beta_1(\omega)$ and $\beta_2(2\omega)$ are respectively the propagation constants of the F and SF modes whose angular frequencies are respectively $\omega$ and $2\omega$.
$k_i$ ans $t_i$ are the coupling and transmission coefficients (with $|k_i|^2+|t_i|^2=1$) between the guides and the resonators, we can write the corresponding coupling matrices \cite{Yariv:00}
\begin{align}
M_i= \left(\begin{array}{cc}
t_i & k_i\\
-k^\ast_i & t_i^\ast\\
\end{array}\right)
\end{align}
where $i=1$ for the F field and $i=2$ for the SH field . The coupling between the two resonators is described by two coefficients $\kappa_i$ and $\tau_i$ (with also $|\kappa_i|^2+|\tau_i|^2=1$) and thus by the matrix 
\begin{align}
P_i= \left(\begin{array}{cc}
\tau_i & \kappa_i\\
-\kappa^\ast_i & \tau_i^\ast\\
\end{array}\right).
\end{align}
We have for instance $(A_T~A_1)^T=M_1(A_{in}~A_4)^T$, $(B_D~B_7)^T=M_2(0~B_6)^T$ and $(A_3~A_5)^T=P_1(A_{2}~A_8)^T$. We consider that the pump is undepleted and thus all the calculations are linear for the F fields. Using the matrices $M_1$ and $P_1$ it is straightforward to calculate the field $A_i(s)$ at each point of the system \cite{Dumeige:06b}. In the case of zinc-blende symmetry materials such as GaP, the second-order nonlinear susceptibility depends on $s$ in the form $\chi^{(2)}(s)=\chi_{0+}^{(2)}e^{j\frac{2s}{R}}+\chi_{0-}^{(2)}e^{-j\frac{2s}{R}}$ \cite{Dumeige:06a, Yang:07}. The SH field is then calculated along each half perimeter of the resonator by integrating the following equation
\begin{equation}
\frac{d\mathcal{B}_{2p+1}}{ds}=A_{2p+1}^2(s_0)\left(\chi_{0+}^{(2)}e^{j\frac{2s}{R}}+\chi_{0-}^{(2)}e^{-j\frac{2s}{R}}\right)e^{j\Delta\beta s}
\end{equation}
for $p\in[0,3]_{\mathbb{N}}$ with $\Delta\beta=\beta_{2}-2\beta_{1}$; $s_0=0$ for $p=\{0,2\}$ and $s_0=L$ for $p=\{1,3\}$ . Defining $\varphi_2=-\beta_2L$, we obtain the expression of the amplitude of the output field on the trough port 
\begin{equation}\label{SHG_calc}
B_T=k_2Ke^{j\varphi_2}\frac{\left[H_2A_1^2+\frac{\kappa_2}{D}\left(t_2^*A_5^2+A_7^2e^{-j\varphi_2}\right)+A_3^2e^{-j\varphi_2}\right]}{1-t_2^*H_2e^{2j\varphi_2}}
\end{equation}
with 
\begin{equation}
D=1-t_2^*\tau_2^*e^{2j\varphi_2},
\end{equation}
and where $H_2(2\omega)$ is the linear transfer function of the right microdisk for the SH field (see Fig. \ref{Fig1})
\begin{equation}
H_2=\frac{\tau_2-t_2^*e^{2j\varphi_2}}{1-t_2^*\tau_2^*e^{2j\varphi_2}}.
\end{equation}
For zinc-blende symmetry materials, the nonlinear coupling is characterized by
\begin{equation}
K=j\left[\chi_{0+}^{(2)}\frac{1-e^{\widetilde{j\Delta\beta}_+L}}{\widetilde{\Delta\beta}_+}+\chi_{0-}^{(2)}\frac{1-e^{j\widetilde{\Delta\beta}_-L}}{\widetilde{\Delta\beta}_-}\right]
\end{equation}
where the effective phase mismatch $\widetilde{\Delta\beta}_{\pm}$ is defined as
\begin{equation}\label{4bar}
\widetilde{\Delta\beta}_{\pm}=\Delta\beta\pm\frac{2}{R}.
\end{equation}
In standard nonlinear photonic devices, the effective index mismatch between the F and SH modes depends only on the material and modal dispersions.  It is independent of the cavity length (or micro-resonator radius). On the contrary, the  phase mismatch compensation offered by the  $\overline{4}$ $\chi^{(2)}$ tensor symmetry, see (Eq. \ref{4bar}), as well as the maximal splitting of coupled resonators both scale with the free spectral range (FSR). Their cumulated effect thus requires small geometries to effectively cancel the effective index mismatch. We apply the previous model to the case of two identical coupled GaP microdisks grown along the $z=[001]$ crystallographic axis and embedded in air as described in Fig. \ref{Disks}. The disks are separated by an air gap $g$, their thickness is $H=136~\mathrm{nm}$ and their radius $R=2.7~\mu\mathrm{m}$.
The properties of the system is obtained from the optical properties of a single resonator. The confinement in the $z$-direction is taken into account using the effective index method (EIM) \cite{Kuo:11}. We used the lowest order guided modes in the $z$-direction both for the F and SH frequencies to optimize the effective nonlinear susceptibility. Due to the $\overline{4}3m$ symmetry of the GaP, the F field is chosen polarized in the $(x,y)$ plan which gives an SH field polarized in the $z$-direction. Thus in the EIM framework, the F field is TE polarized whereas the SH field in TM polarized. The effective propagation constants calculated using this approximation are noted $\gamma_1$ and $\gamma_2$ respectively for the F and SF fields. The propagation constants $\beta_1$ and $\beta_2$ of the two modes are obtained solving the Helmholtz equation for a cylinder of refractive index $\gamma_{1,2}/k_0$ (with $k_0(\omega)$ the free space wavevector) surrounded by air. Note that for both frequencies we have chosen the modes with the lowest radial order to reach a good overlap of the interacting fields and thus a maximal effective non-linear susceptibility \cite{Dumeige:06a}. In this case it is not possible to use the modal phase matching technique to compensate for the dispersion to reach the phase matching condition. Figure \ref{Spectrelin}.a) shows the linear transmission spectra for a single disk on the drop port assuming that the intrinsic and coupling quality (Q) factors are equal to $10^4$. $\nu_F$ is the F mode frequency, the frequency $\nu_{SH}$ of the SH field has been divided by two for the sake of clarity.
Due to the strong material and form dispersions it is not possible to reach the double resonance with a difference in the phase mismatch such as $\Delta m-2=0$ (where $\Delta m=\Delta\beta R$ is the phase mismatch in term of azimutal quantum numbers), consequently $\nu_{F0}$ and $\nu_{SH0}$ the resonant frequencies for the F and the SH modes giving a minimal phase mismatch do not coincide and $\nu_{F0}>\nu_{SH0}/2$. The double resonance $\nu_{F0}=\nu_{SH0}/2$ could be obtain but with $\Delta m-2$ equal to an integer strictly positive which would lead to a null SHG over a circumference of the resonator. By coupling two identical resonators, the F and SH resonances are split by respectively $\Delta\nu_{F}$ and $\Delta\nu_{SH}$, values depending on $g$. If we define the new frequency mismatch between the split resonant frequencies 
\begin{equation}
\delta(g)=\frac{\nu_{SH0}}{2}-\nu_{F0}-\frac{1}{2}\left(\Delta\nu_F(g)+\frac{\Delta\nu_{SH}(g)}{2}\right)
\end{equation}
it is possible to find an air-gap value $g_0$ which makes the two inner F and SH frequencies to match reaching the phase matching as shown in Fig. \ref{Spectrelin}.b). At the same time we have $|\Delta m-2|<1/2$ which insures an efficient SHG over the perimeter of each resonator. To find the value of $g_0$ we used a home-made FDTD code \cite{Jebali:20} to calculate the linear spectra around F and SH frequencies for the coupled structure for several values of $g$.
\begin{figure}[h!]
\centering
\includegraphics[width=\linewidth]{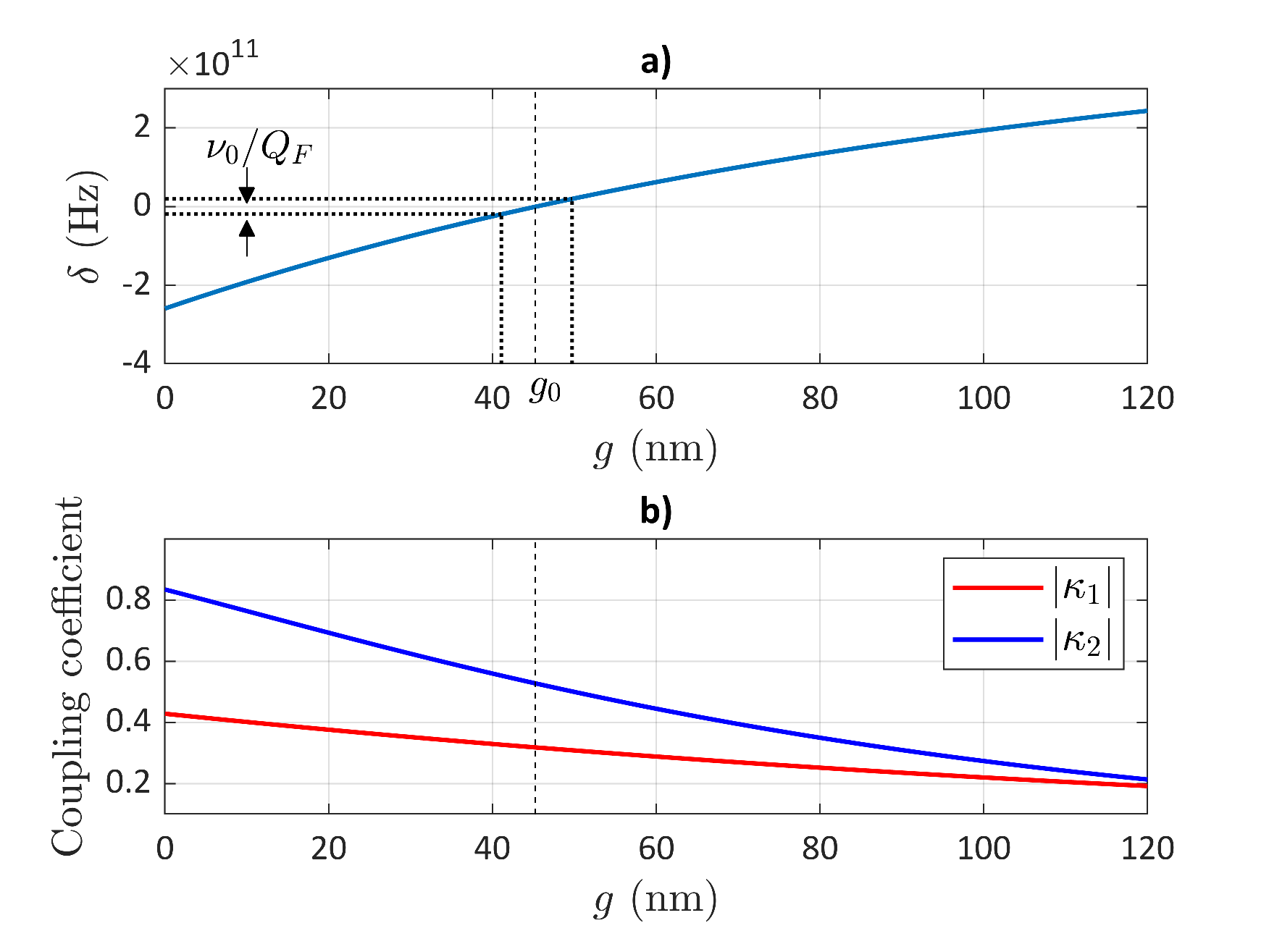}
\caption{a) Nonlinear detuning $\delta(g)$ calculated using a combinations of FDTD simulations and EIM. b) Coupling coefficients $\kappa_1$ and $\kappa_2$ deduced from FDTD simulations. The vertical dashed line gives the air gap value $g_0$ fulfilling the double resonance condition.}
\label{FDTD}
\end{figure}
Thus, the detuning $\delta$ can be calculated varying the air-gap as shown in Fig. \ref{FDTD}.a). Note that to produce the results given Fig. \ref{FDTD}, we interpolated the numerical discrete calculations of $\Delta\nu_F(g)$ and $\Delta\nu_{SH}(g)$ using decaying exponential functions. The corresponding coupling coefficients necessary to calculate the SHG in the coupled mode framework using Eq. (\ref{SHG_calc}) are deduced from 
$|\kappa_1|=\sin{\left(\frac{\pi\Delta\nu_{F}}{\Delta_1}\right)}$ and $|\kappa_2|=\sin{\left(\frac{\pi\Delta\nu_{SH}}{\Delta_2}\right)}$ where $\Delta_1$ an $\Delta_2$ are respectively the FSR for the F and SH fields \cite{Smith:03}, the resulting values are plotted in Fig. \ref{FDTD}.b). For a microdisk, the effective nonlinear susceptibilities is calculated for the single resonator WGM modes of resonant frequencies $\nu_{F0}$ and $\nu_{SH0}$ and azimutal quantum numbers $m_1=17$ and $m_2=36$. Cylindrical coordinates $(r,\theta)$ are defined in Fig. \ref{Disks} and $d_{14}=28.5~\mathrm{pm/V}$ is the nonlinear coefficient of bulk GaP \cite{Guilleme:16}. From the coupling mode theory \cite{Dumeige:06a} we find 
\begin{equation}\label{chi2eff}
\chi_{0\pm}^{(2)}=-\frac{j\gamma_2}{4R^2N^2_{2}}\int_0^Rr^2d_{14}f_{\pm}(r)(\mathcal{H}_r^{SH})^*dr,
\end{equation}
where $N_2$ is the bulk refractive index of GaP at the SH frequency. In expression (\ref{chi2eff}) we have assumed that
\begin{equation}
f_{\pm}(r)=\mathcal{E}_r^{F}\mathcal{E}_{\theta}^{F}\pm j\frac{(\mathcal{E}_r^{F})^2-(\mathcal{E}_{\theta}^{F})^2}{2}.
\end{equation}
The F electric fields can be expressed using $J_{\alpha}$ the first kind Bessel functions of order $\alpha$ 
\begin{equation}
\mathcal{E}_r^{F}=-\frac{a m_1}{r\epsilon_0}\left(\frac{2\pi\nu_{F0}}{\gamma_1 c}\right)^2J_{m_1}(\gamma_1r)
\end{equation}
and 
\begin{equation}
\mathcal{E}_{\theta}^{F}=-\frac{j}{m_1}\frac{d\left(r\mathcal{E}_r^{F}\right)}{dr}.
\end{equation}
Eventually, the SH magnetic field is given by
\begin{equation}
\mathcal{H}_r^{SH}(r)=bJ_{m_2}(\gamma_2r).
\end{equation}
The constants $a$ and $b$ are calculated in order that $\left|\mathcal{A}_i\right|^2$ and $\left|\mathcal{B}_i\right|^2$ correspond to the power flows for the F and SH modes in $\mathrm{W/m}$ \cite{Dumeige:06a}. 
\begin{figure}[h!]
\centering
\includegraphics[width=\linewidth]{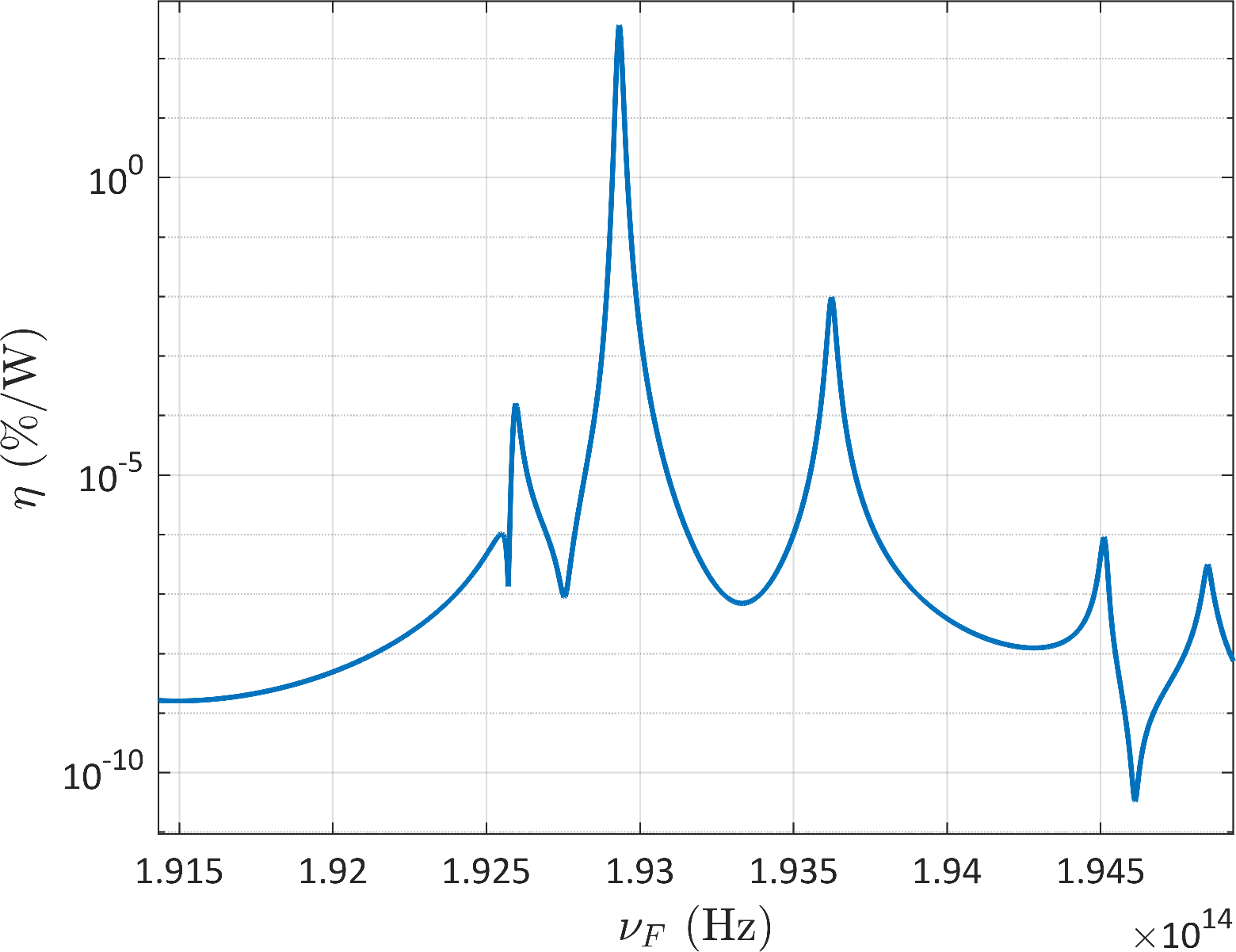}
\caption{Conversion efficiency $\eta(\nu_F)$ calculated for $\delta(g_0)=0$.}
\label{Conversion}
\end{figure}
Figure \ref{Conversion} represents the normalized conversion efficiency $\eta=\frac{|B_T|^2}{|A_{in}|^2\mathcal{P}_{in}}$ (where $\mathcal{P}_{in}$ is the input F power) for the studied structure in the case $g=g_0$. For the doubly resonant frequency we obtain a maximal conversion efficiency of $360\%/\mathrm{W}$ which is consistent with results calculated for small thin microdisks (with same Q-factors) under perfect phase matching \cite{Guilleme:16}. This result shows that even though we obtained the double resonance with two supermodes with different symmetry it does not affect the efficiency of the device. The interest of the proposed method is that the resonator coupling gives an additional free parameter which would allows to use thicker and largest disks which would be less sensitive to fabrication imperfections. As with single doubly resonant resonators, the disk radius must be controlled with an accuracy of the order of $R/Q_F$. In the example presented, this gives $0.5~\mathrm{nm}$. For practical implementation, it will be necessary to control the resonances independently, using, for example, thermal effects \cite{Borghi:19}. Another solution could be to use a single corrugated disk and take advantage of the coupling of co- and counter-propagating modes \cite{Shirpurkar:22}. This would mean that only one resonator would need to be controlled. If we suppose a tolerance of $\nu_{F0}/Q_F$ (with $Q_F$ the loaded Q-factor for the F) for the double resonance in frequency it translates in a 10~nm fabrication tolerance for the air-gap. Note that the relative phase between the two coupling coefficients $\kappa_1$ and $\kappa_2$ has a negligible effect on the conversion efficiency maximum.

We have proposed a new method to reach the phase matching condition in second-order nonlinear processes in integrated optics relying on the use of the frequency splitting due to the coupling of identical resonators. We show how this phase matching scenario could be implemented using GaP microdisks to reach conversion yields such high as $360\%/\mathrm{W}$ with micron-size devices. The resonator coupling could also be used as the additional free parameter required to reach the double resonance condition in single mode nonlinear Fabry-Perot microcavities \cite{Berger:97}. This method could be straightforwardly applied to a chain of three resonators which could be useful to reach higher frequency splittings for a given coupling strength between resonators enabling the compensation of larger phase-mismatch. Finally our approach could be profitably applied to the SHG in nano-resonators \cite{Raineri:02} such as coupled photonic crystal cavities. Indeed, since the splitting strongly depends on the FSR of the cavity, huge detuning as high as several nanometers could be canceled in these ultimately short cavities \cite{Azzini:13}. 

\section*{Funding Information}
This research was supported by the French National Research Agency (ANR) through the JCJC Orpheus project (ANR-17-CE24-0019) and the "France 2030" OFCOC project (ANR-22-PEEL-0005). 

%\section*{Disclosures}
%The authors declare no conflict of interest.
%
%\section*{Data availability}
%Data underlying the results presented in this paper are not publicly available at this time but may
%be obtained from the authors upon reasonable request.

% Bibliography
%\bibliography{sample}

% Full bibliography added automatically for Optics Letters submissions; the following line will simply be ignored if submitting to other journals.
% Note that this extra page will not count against page length
%\bibliographyfullrefs{sample}

%Manual citation list
%\begin{thebibliography}{1}
%\bibitem{Zhang:14}
%Y.~Zhang, S.~Qiao, L.~Sun, Q.~W. Shi, W.~Huang, %L.~Li, and Z.~Yang,
 % \enquote{Photoinduced active terahertz metamaterials with nanostructured
  %vanadium dioxide film deposited by sol-gel method,} Opt. Express \textbf{22},
  %11070--11078 (2014).
%\end{thebibliography}

\clearpage
%\ifarXiv
%    \foreach \x in {1,...,\numbersupplementpages}
%    {
%        %\clearpage
%        \includepdf[pages={\x,{}}]{\supplementfilename}
%    }
%\fi
\includepdf[pages={1}]{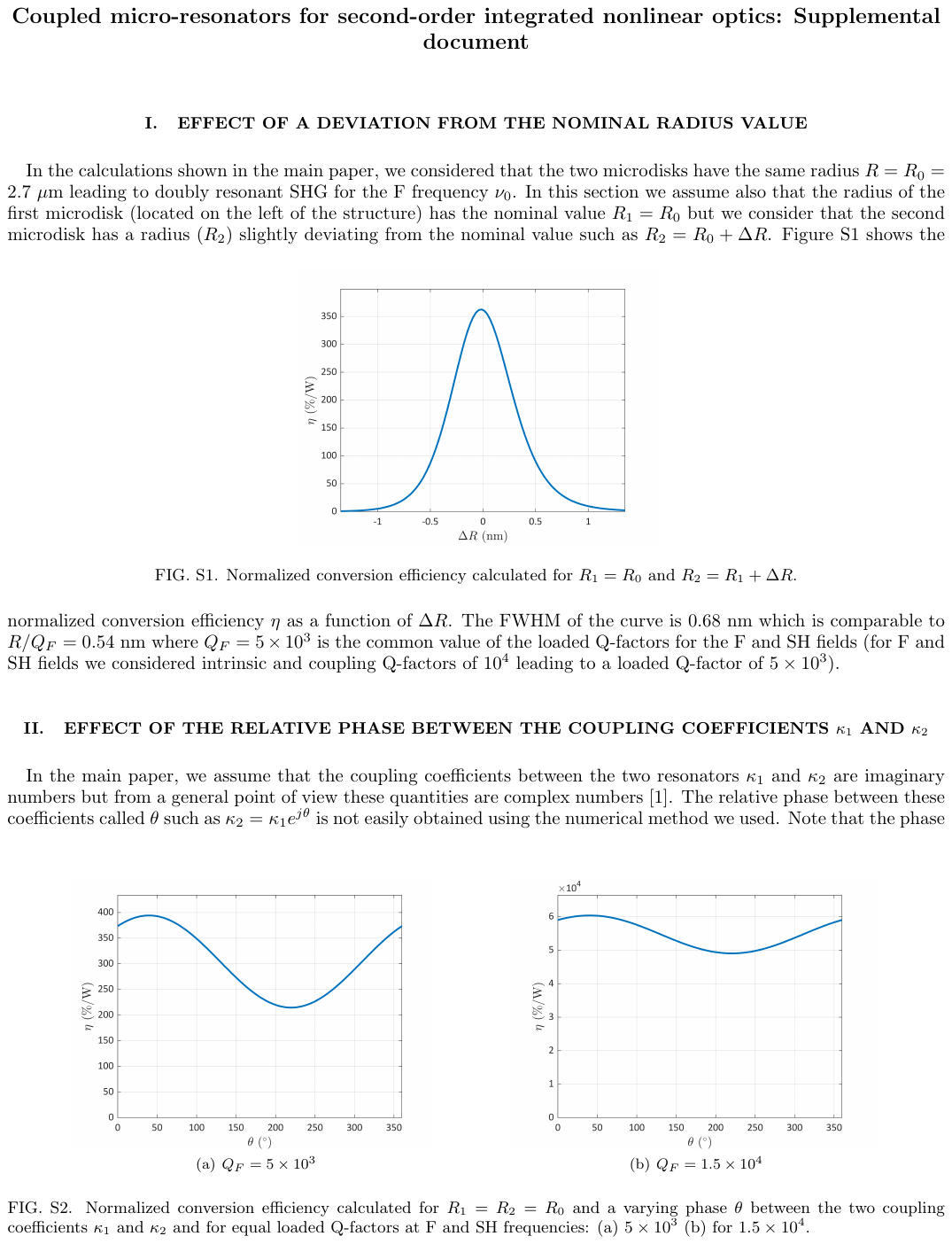}
\clearpage
\includepdf[pages={2}]{Supplementary.pdf}
\clearpage
\includepdf[pages={3}]{Supplementary.pdf}

\end{document}